\documentclass[sigconf]{acmart}
\AtBeginDocument{%
  }

\setcopyright{none}
\copyrightyear{2026}
\acmYear{2026}
\acmConference{ACM}{2026}

\begin{document}

\title{Generative AI User Experience: Developing Human--AI Epistemic
Partnership}

\author{Xiaoming Zhai}
\email{xiaoming.zhai@uga.edu}
\affiliation{%
  \institution{AI4STEM Education Center, University of Georgia}
  \city{Athens}
  \state{GA}
  \country{USA}
}








\renewcommand{\shortauthors}{Zhai}

\begin{abstract}
  Generative AI (GenAI) has rapidly entered education,
yet its user experience is often explained through adoption-oriented
constructs such as usefulness, ease of use, and engagement. We argue
that these constructs are no longer sufficient because systems such as
ChatGPT do not merely support learning tasks but also participate in
knowledge construction. Existing theories cannot explain why GenAI
frequently produces experiences characterized by negotiated authority,
redistributed cognition, and accountability tension. To address this
gap, this paper develops the Human--AI Epistemic Partnership Theory
(HAEPT), explaining the GenAI user experience as a form of epistemic
partnership that features a dynamic negotiation of three interlocking
contracts: \emph{epistemic, agency,} and \emph{accountability}. We argue
that findings on trust, over-reliance, academic integrity, teacher
caution, and relational interaction about GenAI can be reinterpreted as
tensions within these contracts rather than as isolated issues. Instead
of holding a single, stable view of GenAI, users adjust how they relate
to it over time through \emph{calibration cycles}. These repeated
interactions account for why trust and skepticism often coexist and for
how partnership modes describe recurrent configurations of human--AI
collaboration across tasks. To demonstrate the usefulness of HAEPT, we
applied it to analyze the UX of collaborative learning with AI speakers
and AI-facilitated scientific argumentation, illustrating different
contract configurations.
\end{abstract}





\maketitle

\textbf{\hfill\break
}

\section{Introduction}\label{introduction}

Educational technologies have historically been theorized as instruments
that deliver content, structure practice, represent phenomena, or
coordinate interaction (Januszewski \& Molenda, 2013). Even when
learners become emotionally engaged with games or socially connected
through platforms, the technology has typically remained epistemically
subordinate: it is not presumed to \emph{participate} in knowledge
construction as a quasi-interlocutor. Generative artificial intelligence
(GenAI) like ChatGPT changes the phenomenology and the stakes of
educational interaction by producing fluent language that resembles
explanation, critique, and reasoning on demand (Samala et al., 2024). In
classrooms, such outputs function as candidate knowledge claims and
candidate artifacts, entering directly into the epistemic stream of
learning activity---often prior to teacher mediation or peer review
(Memarian \& Doleck, 2023).

Two structural conditions accelerate the educational significance of
this shift. First, much GenAI is widely accessible and used beyond
institutional control, which reduces the ``gatekeeping'' function that
schools historically exerted over instructional technologies. The OECD's
\emph{Digital Education Outlook 2026} underscores that, unlike earlier
waves of education technology, GenAI is often freely accessible,
intuitive, and adopted outside school oversight, and it warns that
outsourcing tasks to GenAI without pedagogical guidance can yield
performance gains without learning gains (OECD, 2026).~ Second, the
policy and regulatory environment remains in flux; iterative releases of
publicly available GenAI are outpacing national regulatory adaptation,
and the absence of regulations leaves data privacy unprotected and
institutions underprepared to validate tools pedagogically and ethically
(UNESCO, 2023).~

The emerging research base suggests that users are already constructing
distinct forms of experience around GenAI. Students primarily used
ChatGPT for brainstorming, summarizing texts, and finding research
articles; they viewed it as useful for simplifying complex information
but less reliable for providing information and supporting classroom
learning, and they expressed concerns about cheating and plagiarism
alongside positive emotions such as curiosity and calmness (Ravšelj et
al., 2025). This user experience is not reducible to convenience such as
immediacy (reassurance and validation), equity (access and perceived
degree value), and integrity (policy ambiguity and group-work
vulnerability), which are central themes of lived experience (Holland \&
Ciachir, 2025).~Moreover, students' experience of these GenAI is
significantly associated with their academic background and personality
(Deng et al., 2025).

Educators' experience shows parallel complexity. For example,
instructors at a U.S. university reported moderate-to-high familiarity
with GenAI concepts but limited direct instructional use; importantly,
trust and distrust were related yet distinct rather than opposite poles
of a single continuum, pointing to the need for calibration rather than
simple persuasion (Lyu et al., 2025).~ Meanwhile, K--12 teachers
reported substantial uncertainty and skepticism, with many teachers
unsure and a quarter reporting that AI tools do more harm than good (Lin
\& Pew Research Center, 2024).~ Despite the mixed perceptions, most
teachers anticipate major shifts to teaching and assessment. Some
believe GenAI will have a major or profound impact and identify
anticipated changes that emphasize learning with AI, higher-order
thinking, ethical values, process-oriented assessment, and the
centrality of face-to-face relational learning (Bower et al., 2024).~

These findings and patterns point to a gap---how to theorize the
epistemic role of GenAI in users' experience of learning and teaching?
Much of the educational technology user experience (UX) literature still
privileges constructs that presuppose tool-like mediation: usability,
engagement, perceived usefulness, and intention to use. Such constructs
remain valuable, but they do not explain why GenAI frequently produces
experiences characterized by negotiated authority (``Should I believe
this?''), redistributed cognition (``Who is doing the thinking?''), and
accountability tension (``Whose work is this---and who is
responsible?'').

To address this gap, we propose the Human--AI epistemic partnership
theory (HAEPT), adapting the educational UX question to become epistemic
and normative. HAEPT proposes that GenAI user experience in education is
best explained as the dynamic negotiation of an epistemic partnership
rather than as the adoption of a tool. HAEPT intends to supply an
explanatory core that can organize---and sometimes
re-interpret---findings that otherwise appear fragmented across
motivation, trust, integrity, and classroom orchestration.

\section{Review of generative AI user
experience}\label{review-of-generative-ai-user-experience}

The empirical literature on GenAI, like ChatGPT, in education is
expanding quickly. Systematic reviews in education synthesize benefits
ranging from accessibility, engagement, and cognitive/emotional support
to concerns about academic integrity, over-reliance, and technostress
(Bahroun et al., 2023; Heung \& Chiu, 2025).~ Moreover, research across
educational sectors highlights opportunities (e.g., individualized
support; teacher workload reduction through automation) while repeatedly
identifying integrity and ethical risks as salient obstacles to
integration.~Within this emerging literature, several UX regularities
are already visible.

\subsection{Use is widespread, but ``use'' does not equal
``trust''}\label{use-is-widespread-but-use-does-not-equal-trust}

Although research has suggested broad uptake and a stable set of common
tasks---brainstorming, summarization, and information finding, a
practicum in the classroom is paired with ambivalence about reliability
and classroom learning value.~These seemingly positive use patterns,
which appear to indicate high acceptance, can often mask epistemic
hesitation and strategic caution. Bouyzourn and Birch (2025) found that
perceived expertise and ethical risk could mostly predict overall trust
in GenAI, followed by concerns about ease of use and transparency. Their
findings also suggest that trust is highly task-contingent, particularly
for complex academic work (e.g., coding).~

Instructor evidence reinforces this distinction. The technical barriers
to AI for teachers are much lower than those for conventional
technologies, resulting in greater access to AI. However, teachers often
show insufficient trust in AI due to various factors, such as a lack of
human characteristics and transparency, which grows their anxieties when
using AI in classroom settings (Nazaretsky et al., 2021). In addition,
Guo et al. (2024b) found that users' familiarity and knowledge of AI may
positively impact UX, thereby increasing their trust without inflating
it. Instructors' limited use for direct instructional tasks in the
presence of conceptual familiarity suggests professional caution shaped
by accountability and role responsibility, not merely by perceived
usefulness.~

\subsection{Experience is relational and affective, not only
instrumental}\label{experience-is-relational-and-affective-not-only-instrumental}

A striking development in measurement research is the move from
``interaction'' to ``relationship.'' Traditional technologies have been
highly instrumental, emphasizing the interactivity that transforms the
learning experience. In contrast, GenAI is usually dialogic, with smart
intelligence mimicking human behavior (e.g., grading students' work).
This feature grants AI the agencies that are usually taken by humans,
thus learners experience AI interaction with some of the same
social-cognitive textures historically associated with human
interaction. That is, students experience more relational and affective
than instrumental. Thus, a robust learner--GenAI relationship is
essential and, sometimes, predictive of learning engagement, perceived
cognitive, and motivational effects (Jin, 2025).~~

These findings matter theoretically because they indicate that GenAI UX
is partly constituted by relational cues---responsiveness,
conversational coherence, and social presence-like dynamics rather than
by interface usability alone. Users do not simply ``operate'' GenAI;
they engage in a patterned dialogue with an entity that can feel like an
interlocutor.

\subsection{Integrity and policy ambiguity are not peripheral; they
are constitutive of
UX}\label{integrity-and-policy-ambiguity-are-not-peripheral-they-are-constitutive-of-ux}

Academic integrity concerns are now a persistent theme in both empirical
studies and reviews of research on GenAI use, revealing how GenAI
influences student behavior and academic honesty. Findings have
emphasized both benefits and risks; thus, the field is calling for
research and policy development that are responsive to this emerging
landscape.~Exemplar studies yield substantial empirical evidence that
implies both practical and theoretical developments. For example, by
examining college students' use of ChatGPT in assessment practices,
Kofinas et al. (2025) suggested that academic integrity in the age of
GenAI is not an external governance issue layered on top of educational
systems; instead, it is embedded in the very experience of assessment.
They found that markers were largely unable to distinguish AI-assisted
submissions from non-assisted ones. Consequently, the presence of GenAI
altered how markers approached the assessment process itself. In other
words, integrity concerns reshaped the lived experience of marking, not
merely its outcomes. This has fundamentally changed both students' and
teachers' experiences regarding integrity.

While integrity is experienced as problematic, it is not merely because
cheating is possible; instead, institutional transparency and policy
play significant roles (Bretag et al., 2019; Eaton, 2023). Research has
found that these institutional infrastructures are often absent, leaving
students uncertain about what is legitimate (McCabe et al., 2012).
Moreover, this uncertainty leaves student groups vulnerable when one
member's ``inappropriate'' use risks collective misconduct. This
ambiguity also causes confusion among teacher users when evaluating
student work produced with emerging technologies (Kasneci et al., 2023).
Thus, it is not surprising that many teachers encounter contradictory
user experiences with generative AI: they acknowledge that GenAI might
harm academic integrity by enabling students to present AI-generated
work as their own, while simultaneously emphasizing that educational
impact depends on pedagogical guidance rather than unstructured
outsourcing (Cotton et al., 2024).~

\subsection{Users adopt ``stances'' toward integration that reflect
moral and cultural
positioning}\label{users-adopt-stances-toward-integration-that-reflect-moral-and-cultural-positioning}

UX is also shaped by how communities narrate what AI should be in
education. These collective narratives, embedded in policy discourse,
professional practice, and public debate, construct shared imaginaries
about AI's appropriate roles, capabilities, and limitations. Such
imaginaries orient users' expectations, shape their trust and
skepticism, and implicitly define the criteria by which AI systems are
judged as effective or ethical. Consequently, UX emerges not solely from
interface features or technical performance, but from the sociocultural
meanings that communities attach to AI and the normative visions they
advance for its place in teaching and learning. Prior research on
technology adoption in education similarly shows that perceptions of
digital tools are mediated by institutional norms, cultural values, and
professional identities, which together influence whether technologies
are framed as empowering resources or disruptive threats to established
pedagogical practices (Dwivedi et al., 2023).

Research has identified distinct viewpoints among students and
professors, including profiles characterized by ethical guardianship,
balanced integration, and convenience-oriented enthusiasm. These
research findings highlighted that perceptions of GenAI (e.g., ChatGPT)
are not merely individual preferences but culturally situated stances
toward legitimacy, fairness, and pedagogical coherence (Tsiani et al.,
2025). Higher education communities show substantial variation in how
students and faculty interpret the benefits, risks, and appropriate uses
of generative AI, particularly around issues such as academic integrity,
intellectual dependency, and the preservation of critical thinking
(Johri et al., 2024). While many students perceive generative AI as a
practical learning aid that can enhance productivity and provide
personalized support, educators often adopt a more cautious stance due
to ethical responsibilities and concerns about assessment validity and
the erosion of disciplinary learning processes (Sah et al., 2025).

Such findings align with the broader observation that debates about AI
integration in education are frequently debates about what education is
for in an AI-rich environment. Discussions surrounding AI often invoke
competing visions of educational purpose---whether emphasizing
efficiency and skill augmentation, the cultivation of independent
reasoning, or the preservation of humanistic forms of inquiry (Selwyn,
2021; Zhai, 2022). In this sense, user experience with AI systems is
inseparable from broader normative debates about the goals of schooling,
the meaning of academic work, and the kinds of intellectual capacities
that education should foster in an increasingly automated society.

Taken together, current evidence suggests that GenAI UX is best
described as a composite experience structured by (a) relational
interaction patterns, (b) epistemic uncertainty and evaluation demands,
and (c) institutional and moral accountability pressures. This composite
is not well captured by one-dimensional models of acceptance.

\section{Existing EdTech user experience
theories}\label{existing-edtech-user-experience-theories}

While existing UX theories have largely accounted for conventional
educational technologies, they remain limited in explicating the
distinctive user experiences emerging from the use of GenAI. First,
technology acceptance models in the TAM and UTAUT traditions have been
highly effective in explaining adoption behavior, particularly through
constructs such as perceived usefulness, perceived ease of use,
performance expectancy, and effort expectancy (Davis et al., 1989;
Venkatesh et al., 2003). Yet their explanatory center typically remains
the user's perception of whether a system helps achieve goals
efficiently. With GenAI, however, ``usefulness'' is entangled with
epistemic validity: outputs can feel useful, fluent, and responsive
while still being inaccurate or misleading (Guo et al., 2025; Wang et
al., 2026). Recent research in higher education shows this tension
clearly. For example, students often report positive attitudes toward
GenAI because of its personalized, immediate support and perceived
usefulness, while simultaneously expressing substantial concern about
factual inaccuracy and the inability of these systems to handle complex
tasks reliably (Chan \& Hu, 2023). Research in computer science and
engineering education found similar outcomes. That is, GenAI (e.g.,
ChatGPT) seems effective in programming tasks but performed less well in
more complex domains that require deeper analytical reasoning and
domain-specific expertise (Waqas et al., 2025). All that being said,
perceived output quality is a significant driver of students' learning
motivation and outcomes, underscoring how UX with GenAI is shaped by the
tension between fluency, utility, and uncertainty rather than by stable
instrumental utility alone (Bai \& Wang, 2025).

Second, cognitive and motivational theories explain learning interaction
with the technology, treating the latter as an instructional medium,
which fails to cover GenAI's increasing role an epistemic actor. For
example, cognitive load theory helps explain why some interfaces reduce
search costs and extraneous processing demands during learning (Sweller,
1988). Motivation theories such as self-determination theory likewise
clarify why learners may experience greater autonomy or competence when
they receive immediate, personalized assistance (Ryan \& Deci, 2000).
However, these frameworks were not designed to explain the
social-epistemic negotiation involved when learners interact with a
system like GenAI that produces authoritative-seeming answers, shapes
belief revision, and may encourage reliance even when its outputs
warrant scrutiny. Recent work on epistemic agency argues that AI affects
not only what people know, but also how they form and revise beliefs,
thereby raising questions that extend beyond cognitive efficiency or
motivational support (Coeckelbergh, 2026).

Third, even sociocultural and activity-theoretic approaches, which are
better suited to contextualizing tools in practices, were largely
developed for artifacts that mediate action rather than generate
domain-relevant explanations and arguments at scale that GenAI attends
to. Wertsch (1998) suggest articulates how sociocultural theories
emphasize the ways that tools shape human action within institutional
and historical contexts, while Engeström (2001) employs activity theory
to similarly examine how tools mediate collective activity systems and
their transformations. With GenAI, however, mediation increasingly
becomes co-production. This shift is now being recognized in educational
scholarship that frames GenAI as reconfiguring epistemic authority and
even functioning as a ``surrogate knower'' or ``innovator,'' thereby
challenging epistemic agency and disrupting the justificatory practices
through which knowledge is traditionally built in classrooms (Jose et
al., 2025; Zhai, 2024). Chen (2025) further argues that GenAI should be
understood as epistemic infrastructure rather than as a neutral tool,
because it reshapes the conditions under which epistemic agency can be
exercised and may influence long-term habit formation in knowledge work.

Fourth, the human--AI teaming literature offers important insights into
trust, coordination, transparency, and shared cognition, but education
introduces additional stakes, including developmental aims such as
building learners' agency, institutional norms such as assessment and
credentialing, and asymmetrical responsibility, insofar as teachers
remain accountable for students and the curriculum. Practices in the
science domain have shown that scientists treat AI as a collaborative
agent rather than a tool, highlighting the role of GenAI in the
epistemic process (Herdiska \& Zhai, 2024; Zhai, 2025). That is,
human--AI teaming through adding an AI teammate can reduce coordination,
communication, and trust, and poor mutual understanding frequently
undermines team performance (Schmutz et al., 2024). To keep the
effectiveness, Endsley (2023) suggests that shared situation awareness,
transparent information displays, explainability, and trust calibration
are essential. More recent frameworks also argue that collaborative
human--AI contexts require new approaches to managing trust over time
and to understanding how responsibilities are distributed across
interaction processes and task phases (McGrath et al., 2025). These
insights are highly relevant to GenAI UX but require a theory that
explicitly integrates epistemic authority and institutional
accountability with learning aims.

Finally, measurement innovation itself signals theoretical
insufficiency. The emergence of epistemic trust measures suggests that
researchers are moving beyond traditional adoption constructs to capture
the distinctive experience of learning with GenAI. For example, the
Epistemic Trust in GenAI for Higher Education Scale (ETGAI-HE)
identifies six dimensions of epistemic trust, including cognitive
evaluation of trustworthiness, interpersonal and contextual influences,
dependability and safety, system predictability and transparency,
performance expectation, and user control and autonomy, but the
validated six-factor model explains 70.8\% of the total variance (Pandey
et al., 2025). This progress in measurement is significant because it
indicates that trust in GenAI is multidimensional and structurally
central to the educational experience. At the same time, the need for
such new measures suggests that existing UX theories do not yet provide
an integrating explanatory account of why these dimensions matter, how
they interact, and how they are enacted in authentic educational
activity.

\section{Human--AI Epistemic Partnership
Theory}\label{humanai-epistemic-partnership-theory}

To fill the gap, the Human-AI epistemic partnership theory
(HAEPT) was proposed, claiming that user experience with GenAI in
education is fundamentally an epistemic partnership. This partnership is
a dynamic, context-sensitive negotiation in which a human and a GenAI
system jointly shape knowledge work (e.g., explanation, ideation,
argumentation, modeling, evaluation) under institutional norms that
allocate authority and responsibility. Rather than conceptualizing GenAI
use simply as tool interaction, HAEPT positions human--AI engagement as
a form of distributed cognition in which knowledge production is
mediated through interactions among human reasoning, technological
affordances, and sociocultural context (Hutchins, 1995; Salomon, 1997).
In this sense, the experiential dimension of GenAI use arises not merely
from interface design but from how users perceive the system's epistemic
role within knowledge-making
processes.

HAEPT begins from an empirical observation in addition to a theoretical
stance. Empirically, learners and teachers frequently behave as though
GenAI can make epistemic contributions (Zhai \& Nehm, 2023). That is,
they consult it, ask for a rationale, request alternatives, and
integrate its proposals into knowledge artifacts. Observational and
survey studies across higher education contexts show that students
commonly use generative AI systems such as ChatGPT for brainstorming,
explanation, and drafting tasks, treating the system as a conversational
partner that can provide feedback or generate alternative perspectives
(Dwivedi et al., 2023; Nyaaba et al., 2024). Systematically, this
interaction resembles a partnership arrangement in which roles and
responsibilities must be negotiated.

Theoretically, HAEPT aligns with scholarship arguing that contemporary
AI systems occupy a novel position within epistemic ecosystems: they are
neither traditional information tools nor autonomous experts, but
generative systems capable of producing plausible knowledge
representations at scale (Bender et al., 2021; Floridi et al., 2018). As
a result, educational use of GenAI requires pedagogical approaches that
support critical discernment, epistemic vigilance, and reflective
engagement rather than passive acceptance of machine-generated outputs
(Zhai et al., 2026). From this perspective, UX with GenAI is inseparable
from broader questions about epistemic authority, responsibility, and
the social organization of knowledge production in digital learning
environments.

\subsection{Epistemic partnership in
education}\label{epistemic-partnership-in-education}

A human-AI epistemic partnership in education can be conceptualized as a
socio-cognitive configuration in which human learners, educators, and AI
jointly construct knowledge under specific epistemic and institutional
constraints. This configuration features three analytically distinct yet
theoretically grounded conditions: First, the interaction must involve
the production, transformation, or evaluation of knowledge claims or
artifacts (e.g., explanations, arguments, solutions, instructional
designs, or feedback), with GenAI functioning as an active contributor
to these epistemic processes. In Human--GenAI contexts, this
construction is increasingly mediated by AI-generated outputs that
externalize and extend human thinking. These outputs---such as generated
explanations or drafts---serve as epistemic artifacts that can be
inspected, critiqued, and iteratively refined (Bereiter \& Scardamalia,
1993). Importantly, GenAI does not merely transmit information but
participates in shaping the form and direction of knowledge
construction, thereby functioning as a cognitive agent that augments and
reorganizes epistemic activity.

Second, the interaction must entail the implicit or explicit
distribution of epistemic roles between humans and GenAI (e.g.,
generator, critic, verifier, or authority), with these roles dynamically
negotiated rather than fixed. Researchers have argued that cognitive
processes in Human-AI partnership are not confined to the human learner
but are distributed across human and artificial agents (Hutchins, 1995;
Salomon, 1997). GenAI systems may assume roles such as content generator
or feedback provider, while humans may act as evaluators, curators, or
meta-level regulators of the interaction. However, unlike traditional
tools, GenAI can simulate roles associated with epistemic authority,
raising questions about trust, over-reliance, and the calibration of
human judgment. Thus, the allocation of epistemic roles is not only
constitutive of how knowledge-building processes are organized (Chi,
2009; Dillenbourg, 1999), but also central to understanding how agency
and control are negotiated in AI-mediated learning environments.

Third, the interaction must entail normative consequences, particularly
regarding authorship, accountability, and the legitimacy of knowledge
claims in AI-supported contexts. Educational systems continue to assign
value to processes of knowledge production, justification, and
individual contribution, embedding learning within broader structures of
assessment, credentialing, and professional responsibility (Gee, 2000).
In Human--GenAI partnerships, these normative dimensions become more
complex because AI\textquotesingle s involvement challenges conventional
assumptions about authorship and intellectual ownership. Epistemic
actions---such as submitting AI-assisted work or relying on generated
feedback---are evaluated against institutional norms that define what
counts as valid knowledge and legitimate participation. Consequently,
Human--AI epistemic partnerships are inherently normative, as they
intersect with evolving expectations about transparency, responsibility,
and the appropriate use of AI in educational practice (Crippen et al.,
2026).

Taken together, these three conditions---knowledge production, role
distribution, and normative consequence---are not arbitrary but reflect
foundational concerns in educational research regarding how cognition is
mediated, distributed, and evaluated. In particular, they align with
longstanding inquiries into how technologies reconfigure the division of
cognitive labor and reshape the processes through which learners
construct and validate knowledge (Salomon, 1997). Here, we deliberately
distinguish epistemic partnership from mere interactivity. Many
educational technologies are interactive, but they are not typically
experienced as ``contributors'' that propose expansive content and
reasoning on demand. GenAI differs in that it can produce extended
explanations, arguments, and creative proposals that simulate expert
discourse, thereby blurring the perceived boundary between tool and
collaborator (Kasneci et al., 2023).

\subsection{The three-contract
architecture}\label{the-three-contract-architecture}

We conceptualize HAEPT as a negotiation among three
interlocking ``contracts.'' We use the term ``contracts'' deliberately
to emphasize that these relationships are not merely guided by abstract
principles or unilateral rules, but by mutually constituted expectations
that carry implications for rights, responsibilities, and
accountability. The contract metaphor reflects sociotechnical research
showing that human--technology relationships often rely on tacit
expectations regarding trust, responsibility, and legitimacy (Floridi et
al., 2018). It also highlights that these expectations are shaped
through interaction and can be revised, contested, or even broken,
features that are not fully captured by terms like ``rules'' or
``guidelines.'' These are not formal documents; they are lived, often
implicit, normative arrangements that become visible in moments of
uncertainty, error, evaluation, or
conflict.

\subsubsection{The Epistemic
Contract}\label{the-epistemic-contract}

The Epistemic Contract refers to how users decide what to trust and what
counts as valid knowledge when interacting with AI. It addresses
questions such as: What role is ChatGPT playing in this task---is it a
source of information, a helper, a critic, or simply a text generator?
What kinds of evidence or explanations are expected? How should
uncertainty or possible errors be treated?

This contract is particularly important in the context of GenAI because
these systems produce fluent and confident responses regardless of their
accuracy. This can lead users to overestimate the system's authority or
rely on it uncritically, increasing the risk of automation bias (Bender
et al., 2021). Prior research has described this phenomenon as treating
AI as a ``surrogate knower,'' in which learners may accept outputs
without verifying the evidence or engaging in reflection (Kasneci et
al., 2023). Empirical studies further highlight this tension: while
students often find generative AI tools helpful and engaging, they also
express concerns about their reliability and accuracy for academic
purposes (Deng et al., 2025). In this sense, the Epistemic Contract is
where issues of fluency, trust, and learner judgment come together. It
shapes whether learners treat AI-generated responses as tentative inputs
requiring evaluation or as authoritative answers that can be accepted
without further scrutiny.

\subsubsection{The Agency Contract}\label{the-agency-contract}

The Agency Contract indicates how thinking and decision-making
responsibilities are shared between the user and the AI. It focuses on
how cognitive and metacognitive tasks are distributed, including who
generates ideas, who evaluates their quality, who monitors progress, and
who determines when a task is complete. It raises key questions such as:
Who is actually doing the thinking in this interaction? Am I outsourcing
core cognitive work or using AI to support and extend my reasoning? Who
controls the direction of inquiry, and who is responsible for verifying
the accuracy and quality of the output? In this sense, the Agency
Contract makes explicit the often implicit division of intellectual
labor between human and AI, highlighting how different configurations of
this division can shape both the process and outcomes of learning.

This contract helps explain why simply ``using'' AI does not
automatically lead to meaningful learning. The OECD (2023) distinguishes
between improved performance (e.g., completing tasks more quickly) and
actual learning gains, noting that heavy reliance on generative AI can
increase productivity without improving understanding if users disengage
cognitively. The Agency Contract describes how this gap can emerge. When
AI takes over both idea generation and evaluation, learners may become
less actively involved in the thinking process. In contrast, when AI is
used as a partner that encourages questioning, critique, and reflection,
learners are more likely to stay cognitively engaged.

Research in science education supports this distinction. Studies of
AI-supported dialogic learning environments show that well-designed
conversational agents can promote perspective-taking, argumentation, and
reflective reasoning, particularly in discussions of complex issues.
These findings suggest that the educational value of AI depends not
simply on whether it is used, but on how responsibility for thinking is
distributed within the interaction (Guo et al., 2024a; Watts et al.,
2025).

\subsubsection{The Accountability
Contract}\label{the-accountability-contract}

The \emph{Accountability Contract} refers to how authorship,
responsibility, and ethical expectations are defined when using GenAI,
particularly in contexts where the consequences of use matter. It
focuses on how credit is assigned, what kinds of disclosures are
required, and who is held responsible for the outcomes produced with AI
support. Key questions include: Who ``owns'' the work when AI
contributes to its creation? To what extent should AI use be made
visible to others, such as instructors or collaborators? Who is
accountable if the output is inaccurate, biased, or violates academic or
professional standards? More broadly, how do institutional policies,
assessment practices, and disciplinary norms determine what counts as
legitimate use? By making these issues explicit, the Accountability
Contract highlights that AI use is not only a technical or cognitive
matter, but also a social and ethical one, shaped by expectations of
integrity, transparency, and responsibility.

This contract is central to the UX in educational settings because these
environments attach high stakes to academic integrity, authorship, and
the value of credentials. As a result, even small ambiguities about the
use of GenAI can create significant uncertainty for both students and
instructors. Recent analyses describe generative AI as a major
disruption to established norms of originality and authorship, requiring
institutions to rethink policies, assessment design, and instructional
practices (Gao et al., 2025). Student perspectives further illustrate
how this disruption is experienced in practice: learners often raise
concerns about fairness (e.g., unequal access or inconsistent rules),
transparency (e.g., unclear expectations for disclosure), and
vulnerability in collaborative work (e.g., uneven contributions when AI
is used differently across group members). At the same time, instructors
tend to approach GenAI more cautiously, not simply due to skepticism
about the technology, but because accountability is concentrated in
their role. They are responsible for upholding academic standards,
ensuring fair assessment, and protecting the credibility of credentials,
all of which heighten the perceived risks of AI use. This asymmetry
helps explain why educators may impose stricter boundaries or require
clearer justification for AI-assisted work, and more broadly, why
judgments about the legitimacy of AI use are deeply shaped by
accountability considerations.

In all, the three contracts showed a significant difference between
traditional EdTech and GenAI (see Table 1). Compared with traditional
EdTech, GenAI reconfigures all three contracts, making previously stable
expectations more fluid and negotiable. In traditional EdTech, the
\emph{Epistemic Contract} is anchored in curated, institutionally
validated knowledge, allowing learners to rely on relatively stable
sources of authority with a limited need for ongoing verification. The
\emph{Agency Contract} is similarly structured, with technology
primarily supporting rather than performing cognitive work, leaving
learners responsible for core processes of thinking and evaluation. The
\emph{Accountability Contract}, in turn, remains largely human-centered,
with clear norms around authorship, responsibility, and academic
integrity. By contrast, GenAI destabilizes these arrangements. Its
ability to generate fluent yet uncertain outputs shifts the
\emph{Epistemic Contract} toward continuous calibration of trust and
evidence. At the same time, the \emph{Agency Contract} becomes more
fluid, as cognitive responsibilities can be redistributed between human
and AI in ways that may either support or undermine meaningful learning.
Finally, the \emph{Accountability Contract} becomes more complex and
contested, as questions of authorship, disclosure, and responsibility
are no longer self-evident but must be actively negotiated within
evolving institutional and social norms.

\begin{table*}[t]
\caption{Comparison between traditional Educational Technology and GenAI across three contracts}
\label{tab:three-contract-comparison}
\centering
\small
\setlength{\tabcolsep}{4pt}
\begin{tabular}{p{0.18\textwidth} p{0.37\textwidth} p{0.37\textwidth}}
\toprule
\textbf{Contract} & \textbf{Traditional Educational Technology} & \textbf{GenAI} \\
\midrule

\textbf{Epistemic Contract} \newline
\emph{(What counts as knowledge? What to trust?)} &
Knowledge is pre-curated, stable, and institutionally validated (e.g., textbooks, LMS content, simulations). Trust is largely delegated to external authorities such as teachers, publishers, and curriculum designers. Uncertainty is minimized and bounded. &
Knowledge is dynamically generated, probabilistic, and potentially fallible. Trust must be actively evaluated by the user. Outputs are fluent but may be inaccurate, requiring verification and critical evaluation. Risks include automation bias and treating AI as a surrogate knower. \\

\addlinespace[4pt]

\textbf{Agency Contract} \newline
\emph{(Who does the thinking?)} &
Cognitive roles are clearly structured: learners engage in tasks designed by instructors, while technology supports delivery, practice, or visualization. Thinking remains primarily human-driven, with tools scaffolding specific processes. &
Cognitive labor is fluid and negotiable. AI can generate ideas, evaluate responses, and guide inquiry, potentially taking over core thinking processes. Learning depends on whether users outsource cognition or use AI as a partner for reflection and critique. Risks arise when agency shifts too heavily to AI. \\

\addlinespace[4pt]

\textbf{Accountability Contract} \newline
\emph{(Who is responsible?)} &
Authorship and responsibility are relatively clear and human-centered. Outputs are typically produced by students, with tools playing a transparent, supportive role. Institutional norms for assessment and integrity are well established. &
Authorship becomes ambiguous and distributed between human and AI. New norms are required for disclosure, credit, and ethical use. Responsibility for errors, bias, or misconduct is contested and context-dependent, raising concerns about fairness, transparency, and academic integrity. \\

\bottomrule
\end{tabular}
\end{table*}

\subsection{Calibration cycles: how partnership UX evolves over
time}\label{calibration-cycles-how-partnership-ux-evolves-over-time}

HAEPT emphasizes that epistemic partnership with GenAI is not fixed but
continually evolving. Rather than holding a single, stable view of
GenAI, users adjust their relationship to it over time through repeated
interactions. This process can be understood as a series of
``calibration cycles,'' in which users refine their expectations and
behaviors based on experience. A typical cycle includes: (a)
encountering an AI-generated response, (b) evaluating it---either
implicitly or explicitly---by checking, cross-referencing, or accepting
it, (c) forming an affective reaction such as confidence, relief, or
doubt, and (d) adjusting their approach moving forward, such as trusting
the AI more, relying on it less, or changing how they disclose its use.
Over time, these cycles shape how users position AI in terms of trust,
responsibility, and use. This process is similar to the concept of
calibrated trust in human--automation interaction, in which users
gradually adjust their reliance on automated systems based on their
perceived reliability and task demands (Lee \& See, 2004).

One example of the evolving process is reflected in users' seemingly
contradictory views of GenAI, which are better understood as part of an
ongoing calibration process rather than as fixed attitudes. Empirical
studies consistently show that students report high satisfaction with
generative AI tools due to their usability and perceived quality of
responses, while simultaneously expressing concerns about the accuracy,
reliability, and trustworthiness of AI-generated outputs (Lund et al.,
2026; Ng et al., 2025). Similarly, instructors' trust in AI varies by
context, task, and perceived risk, underscoring the importance of
teaching strategies that encourage the critical evaluation of AI
outputs. This variability and evolution are accounted for by users'
epistemic trust, which is influenced by factors such as transparency,
predictability, and user control. Within HAEPT, these factors serve as
mechanisms shaping how users continuously adjust and renegotiate
epistemic authority, the distribution of cognitive work, and
accountability in human--AI interactions.~

\subsection{Partnership modes: recurrent configurations of the three
contracts}\label{partnership-modes-recurrent-configurations-of-the-three-contracts}

To make the idea of contract negotiation more concrete and observable,
we introduce the concept of \emph{partnership modes} in HAEPT.
Partnership modes refer to relatively stable patterns in how the three
contracts---epistemic, agency, and accountability---are configured
during a particular task. This approach is grounded in prior human--AI
interaction research, which shows that people adopt different
``epistemic relationships'' with AI systems depending on context,
perceived reliability, and purpose of use. In practice, users do not
renegotiate each contract from scratch in every interaction. Instead,
they tend to settle into recognizable ways of working with GenAI (e.g.,
using it as a tool, a collaborator, or an authority), at least within a
given context or activity. These recurring patterns emerge because users
rely on prior experience, task demands, time constraints, and
institutional expectations to guide their interactions. As a result,
partnership modes function as ``default configurations'' that simplify
decision-making during use, even though they remain adjustable over
time. This helps explain why interactions with AI can feel both stable
(within a task) and variable (across contexts or users).

What makes these modes educationally important is that they render
differences in user experience more systematic, interpretable, and
actionable. Rather than attributing variation to vague differences in
attitudes or preferences (e.g., ``some students like AI more than
others''), the framework shows that these differences reflect underlying
configurations of trust, responsibility, and cognitive engagement. For
example, two students may both report frequent AI use, but one may be
operating in a mode of instrumental reliance while the other engages in
co-agency collaboration---leading to very different learning outcomes.
By making these distinctions visible, partnership modes allow
researchers to analyze patterns of use more precisely, link them to
learning processes and outcomes, and develop more nuanced measures of AI
engagement.

Moreover, this framework enables these patterns to be intentionally
shaped through design and pedagogy. Educators can identify which
partnership modes are emerging in their classrooms, evaluate their
alignment with instructional goals, and design interventions to
encourage more productive configurations. For example, prompts,
assignments, and assessment criteria can be structured to promote
verification, reflection, and dialogue, thereby shifting students away
from authority displacement and toward co-agency collaboration.
Similarly, clear policies and disclosure norms can stabilize the
Accountability Contract, reduce ambiguity, and support responsible use.
In this way, partnership modes serve not only as an analytic lens but
also as a practical design tool for guiding more effective and equitable
human--AI interactions in education.

\section{Analytic vignettes: how contracts surface in everyday
educational
moments}\label{analytic-vignettes-how-contracts-surface-in-everyday-educational-moments}

To illustrate HAEPT's explanatory power, consider three analytic
vignettes that are not presented as new empirical data but as
theoretically informed composites consistent with patterns documented in
the literature.

\subsection{Vignette 1. Collaborative learning with GenAI
speakers}\label{vignette-1.-collaborative-learning-with-genai-speakers}

Lee et al. (2023) introduced the CLAIS (Collaborative Learning with
GenAI speakers) system represents an instructional use case in which a
GenAI speaker is embedded as a peer within small-group collaborative
learning, specifically structured through the Jigsaw model. In this
setting, 3--4 pre-service teachers work together with a GenAI speaker
that can respond to spoken prompts, explain content (e.g., learning
theories), and pose questions. The activity proceeds through
expert-group and home-group phases, in which both human students and the
GenAI take turns explaining assigned content and contributing to the
group\textquotesingle s understanding. The GenAI speaker participates
through natural language interaction, offering explanations and answers
based on pre-programmed knowledge, while the instructor facilitates the
process and intervenes when technical issues arise. In this sense, the
use case operationalizes a classroom where GenAI is not merely a support
tool but a conversational participant in collaborative knowledge
construction.

From the perspective of HAEPT, we interpret this use case as a concrete
instantiation of a human--AI epistemic partnership in which knowledge
work is jointly produced through distributed interaction among learners,
AI, and instructional structures. The CLAIS system clearly satisfies the
three defining conditions of epistemic partnership: it involves active
knowledge production (through explanation and problem-solving), dynamic
role distribution (AI as explainer and responder; humans as interpreters
and collaborators), and implicit normative consequences (e.g., peer
evaluation of AI and human contributions). What becomes analytically
significant is how the three contracts---epistemic, agency, and
accountability---are configured in this specific design.

The \emph{epistemic contract} in CLAIS is relatively stable,
characterized by high perceived reliability and limited uncertainty.
Because the GenAI speaker is trained on curated textbook knowledge,
learners tend to treat its outputs as correct, consistent, and
trustworthy, as reflected in their high ratings of accuracy and
reliability. We see this as a partnership mode where GenAI occupies a
quasi-authoritative peer role: it is not formally a teacher, yet it
functions as a dependable source of explanation within the group. The
user experience signature here is one of low epistemic friction:
learners can quickly access and accept explanations without extensive
verification. However, if this epistemic configuration becomes dominant,
it might risk the over-trusting fluency. Learners may not engage in
critical evaluation or epistemic questioning, especially because the
system does not expose uncertainty or alternative interpretations.

The \emph{agency contract} reflects a structured but shifting
distribution of cognitive labor. The Jigsaw design ensures that students
remain responsible for explaining and integrating knowledge, while the
GenAI contributes explanations and answers on demand. In our view, this
produces a partnership mode of guided co-agency, where GenAI supports
and augments human thinking without fully replacing it. The user
experience signature is enhanced efficiency and interactional flow:
students report smoother collaboration and reduced effort in accessing
information. At the same time, we notice that GenAI explanations can
become the model for group discourse, subtly repositioning students as
recipients rather than generators of ideas. If this configuration
dominates, the key educational risk is cognitive offloading, in which
learners increasingly rely on GenAI for core reasoning processes,
resulting in performance gains without corresponding conceptual
understanding.

The \emph{accountability contract} in this use case remains implicit and
under-articulated. Although students evaluate GenAI as a peer, the
system does not explicitly address authorship, responsibility, or
disclosure of GenAI contributions. Interestingly, students apply
different evaluative standards to human and AI peers, being more
critical of GenAI while uniformly positive toward human peers,
suggesting that GenAI is perceived as accountable in performance but not
embedded in social or ethical norms. The user experience signature is,
therefore, a form of normative ambiguity: learners interact with GenAI
as a contributor without clear expectations regarding responsibility for
its outputs. If this contract becomes dominant, it risks diffused
accountability, in which students may incorporate AI-generated knowledge
without fully owning or justifying it, potentially undermining norms of
authorship and academic integrity.

Taken together, we characterize the dominant partnership mode in CLAIS
as structured co-agency with epistemic stabilization and weak
accountability. This configuration produces an engaging and efficient
collaborative experience, demonstrating the feasibility of human--AI
co-participation in the construction of classroom knowledge. However,
from a HAEPT perspective, its limitations are equally instructive: the
design stabilizes trust and participation but does not sufficiently
challenge learners to negotiate epistemic authority, retain cognitive
ownership, or clarify responsibility. These tensions suggest that future
designs should intentionally rebalance the three contracts---introducing
epistemic uncertainty, reinforcing human agency in evaluation, and
making accountability explicit---to support more robust and
educationally productive human--AI epistemic partnerships.

\begin{figure*}[t]
  \centering
  \includegraphics[width=\textwidth]{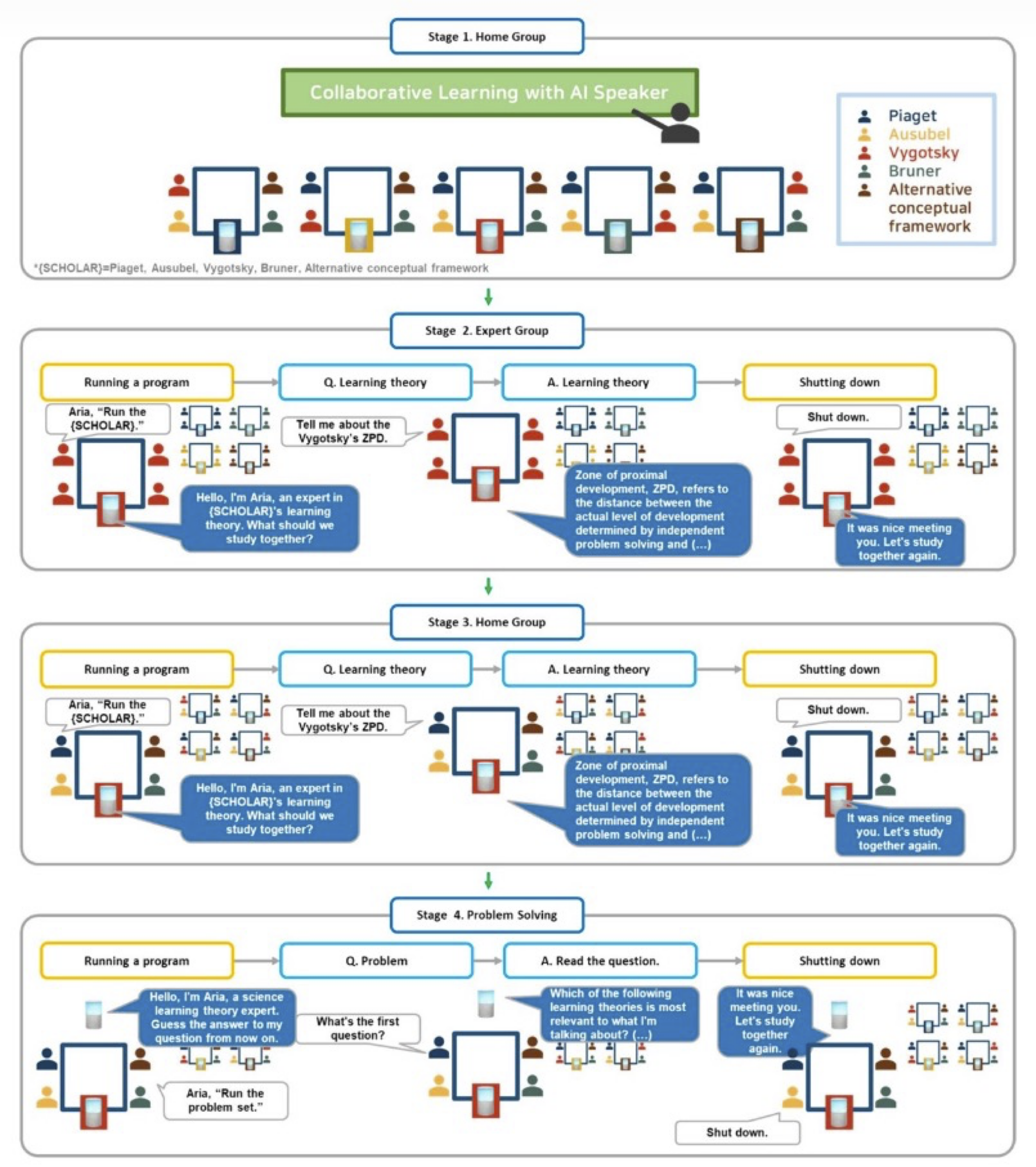}
  \caption{Flow chart of the CLAIS-Jigsaw system (adapted from Lee and Zhai, 2025).}
  \label{fig:clais-jigsaw}
\end{figure*}
Figure 1. Flow chart of the CLAIS-Jigsaw system (adopted from Lee and
Zhai, 2025)

\subsection{Vignette 2. GenAI facilitates live scientific
argumentation}\label{vignette-2.-genai-facilitates-live-scientific-argumentation}

Kleiman et al. (2025, November) developed a GenAI multi-Agent system,
named ArgueAgent, which orchestrates and actively participates in live
scientific argumentation practices. In this system, students first
produce individual explanations or visual models of scientific
phenomena. The GenAI multi-agent system evaluates these responses and
algorithmically pairs students with differing ideas to stimulate
productive disagreement. During subsequent discussion, students are
expected to articulate claims, justify reasoning, critique peers, and
respond to counterarguments. When the argumentation process stalls---due
to lack of skill, premature consensus, or off-track
discussion---ArgueAgent intervenes by taking on roles such as
facilitator (prompting participation), mediator (clarifying positions),
and challenger (posing counterarguments or probing questions). In this
sense, the system does not merely provide content but dynamically shapes
the structure and quality of epistemic interaction in real time.

From the perspective of HAEPT, we interpret ArgueAgent as a more
interventionist and process-oriented epistemic partnership compared to
CLAIS (Lee \& Zhai, 2025). Here, the AI is not primarily a source of
knowledge but an active regulator of epistemic processes---guiding the
generation, evaluation, and contestation of knowledge claims. The system
clearly satisfies the three defining conditions of epistemic
partnership: it engages directly in knowledge production (through
prompting and critique), dynamically redistributes epistemic roles
(e.g., AI as facilitator/mediator/challenger; students as arguers), and
introduces normative consequences (e.g., expectations for participation,
justification, and critique). What becomes central, then, is how the
three contracts are configured to shape users' experience within this
more dialogically intensive environment.

The \emph{epistemic contract} in ArgueAgent is characterized by
productive destabilization of knowledge claims. Unlike CLAIS, where
GenAI outputs are perceived as reliable answers, here GenAI deliberately
introduces tension---by pairing conflicting ideas and challenging
students' reasoning. We see this as a partnership mode in which GenAI
functions as an epistemic provocateur rather than an authority. The user
experience signatures epistemic friction: students are pushed to
justify, defend, and revise their ideas in response to both peers and
GenAI interventions. The activities were configured to highly align with
disciplinary practices in science, where argumentation and critique are
central. However, if this contract becomes overly dominant, there is a
risk of over-reliance on AI-generated interventions. Students may begin
to treat GenAI challenges as the primary standard for evaluation, rather
than developing their own criteria for assessing evidence and arguments.
In this sense, epistemic vigilance may shift from internally regulated
to externally triggered.

The \emph{agency contract} reflects a distributed but AI-orchestrated
division of cognitive labor. ArgueAgent takes responsibility for
structuring interaction (e.g., pairing students, initiating prompts,
sustaining dialogue), while students are responsible for generating and
defending knowledge claims. We interpret this as a partnership mode of
orchestrated co-agency, where GenAI governs the process of thinking
rather than the content itself. The user experience signatures guided
engagement: students are kept actively involved through continuous
prompts and interventions, reducing the likelihood of disengagement or
superficial consensus. At the same time, this strong orchestration
introduces a subtle shift in control, in which AI determines when and
how argumentation proceeds. If dominant, GenAI may risk the process
dependency, where students rely on AI to sustain productive discourse
and may struggle to self-regulate argumentation in its absence. The
development of metacognitive and dialogic skills may thus be constrained
if agency is not gradually rebalanced.

The \emph{accountability contract} in this use case is emergent but
still ambiguous. On one hand, the system implicitly enforces norms of
participation, justification, and critique, thereby strengthening
accountability for epistemic engagement. Students are expected to
contribute and respond, and GenAI interventions make disengagement more
visible. On the other hand, the role of GenAI in shaping argumentation
raises questions about responsibility: who is accountable for the
direction and quality of the discussion---the students or the GenAI that
mediates it? The user experience features a distributed accountability
with hidden governance: students experience themselves as responsible
participants, yet the GenAI exerts significant influence over the
interactional structure. If this contract becomes dominant without
explicit clarification, it can risk blurred epistemic ownership, where
students may attribute the evolution of ideas or the quality of
discourse to the system rather than to their own collective reasoning.

Taken together, we characterize the dominant partnership mode in
ArgueAgent as AI-orchestrated epistemic engagement with productive
tension but latent dependency, which contrasts meaningfully with the
structured co-agency with epistemic stabilization observed in CLAIS (see
Table 2). Whereas CLAIS stabilizes the epistemic contract by positioning
AI as a reliable knowledge contributor, ArgueAgent deliberately
destabilizes it to provoke critique and argumentation. Similarly, while
CLAIS maintains a more balanced---though still shifting---agency
distribution through structured roles in Jigsaw learning, ArgueAgent
concentrates greater control in the AI by orchestrating the flow and
quality of discourse. In terms of accountability, both systems exhibit
ambiguity, but in different forms: CLAIS leaves authorship and
responsibility underspecified in knowledge production, whereas
ArgueAgent introduces accountability for participation and reasoning
while obscuring who governs the epistemic process itself. From a HAEPT
perspective, these differences highlight two distinct trajectories of
human--AI partnership: one that risks over-stabilizing trust and
encouraging passive acceptance (i.e., CLAIS), and another that risks
over-centralizing epistemic regulation and fostering process dependency
(i.e., ArgueAgent). Therefore, we argue that future designs should seek
a balance between these modes---combining ArgueAgent\textquotesingle s
epistemic challenge and engagement with CLAIS\textquotesingle s
structured participation and clearer role distribution---while
intentionally redistributing authority, agency, and accountability back
to learners over time.

\begin{table*}[t]
\caption{Comparison between CLAIS and ArgueAgent}
\label{tab:clais-argueagent}
\centering
\small
\setlength{\tabcolsep}{4pt}
\begin{tabular}{p{0.15\textwidth} p{0.26\textwidth} p{0.26\textwidth} p{0.23\textwidth}}
\toprule
\textbf{Contract} & \textbf{CLAIS (AI Speaker in Collaborative Learning)} & \textbf{ArgueAgent (GenAI for Scientific Argumentation)} & \textbf{Key Difference} \\
\midrule

\textbf{Epistemic} &
Stabilized, high-trust knowledge environment. AI provides reliable, pre-curated explanations. User experience: low epistemic friction, smooth and confident use. Risk: over-trust and reduced critical evaluation; AI may become a surrogate knower. &
Destabilized, contestable knowledge environment. AI provokes critique and conflicting ideas. User experience: epistemic friction through continuous justification and revision. Risk: overreliance on AI critique; epistemic standards may shift externally. &
CLAIS stabilizes knowledge and trust, while ArgueAgent unsettles knowledge to promote critique. \\

\addlinespace[4pt]

\textbf{Agency} &
Structured co-agency. Humans lead explanation and integration; AI supports participation. User experience: efficient, smoother interaction. Risk: cognitive offloading and reduced generative thinking. &
AI-orchestrated co-agency. AI structures pairing and discourse; humans argue within it. User experience: guided engagement and sustained high interaction. Risk: process dependency and weakened self-regulation. &
CLAIS distributes cognition with human leadership; ArgueAgent places process control partly in AI. \\

\addlinespace[4pt]

\textbf{Accountability} &
Implicit and underspecified. AI is treated as a peer but not clearly accountable. User experience: normative ambiguity. Risk: diffuse authorship and weak ownership of knowledge claims. &
Emergent but hidden governance. AI enforces participation and shapes discourse. User experience: distributed accountability with hidden control. Risk: blurred epistemic ownership and responsibility for process. &
CLAIS lacks clear accountability structures; ArgueAgent embeds accountability but obscures who governs it. \\

\addlinespace[4pt]

\textbf{Overall Partnership Mode} &
Structured co-agency with epistemic stabilization and weak accountability. &
AI-orchestrated epistemic engagement with productive tension but latent dependency. &
CLAIS emphasizes stability and support; ArgueAgent emphasizes tension and co-construction. \\

\bottomrule
\end{tabular}
\end{table*}

\section{Discussion}\label{discussion}

Although GenAI has become part of educational life faster than most
earlier technologies, its uptake has not produced a simple story of
acceptance. Students use GenAI widely for brainstorming, summarizing,
searching, and drafting, yet they do not uniformly trust what these
systems produce (Chan \& Hu, 2023; Ravšelj et al., 2025). Instructors
show a similarly mixed stance: many are familiar with GenAI concepts,
but direct instructional use remains limited, and trust often coexists
with distrust rather than replacing it (Lyu et al., 2025). Research adds
further depth by showing that students experience GenAI through themes
such as immediacy, reassurance, equity, and integrity, while teachers
experience it through concerns about assessment validity, fairness, and
policy ambiguity(Cotton et al., 2024; Holland \& Ciachir, 2025). In
other words, users know that GenAI is not experienced as ``just another
tool,'' but it has been less clear how to explain these tensions within
one coherent framework.

We argue that the central issue is not simply whether GenAI is useful,
easy to use, or even engaging. The deeper issue is that GenAI now enters
the space of knowledge work itself. It proposes explanations, drafts
arguments, critiques ideas, and sometimes appears to reason. That
changes the user experience in a fundamental way. Instead of only asking
whether the system helps complete a task, learners and teachers are also
asking: Should I believe this? Who is doing the thinking here? Whose
work is this? Existing acceptance models, such as TAM and UTAUT, remain
valuable for explaining adoption and perceived usefulness (Davis et al.,
1989; Venkatesh et al., 2003), but they are not built to explain these
more epistemic questions. By introducing epistemic partnership as the
unit of analysis, this paper adds a concept that better fits the
realities of GenAI use in education, extending recent work that
describes GenAI as a ``surrogate knower'' or as a new form of epistemic
infrastructure (Kasneci et al., 2023; Chen, 2025). Our argument is that
the user experience of GenAI is not only relational or conversational;
it is epistemic as well. Users are not merely interacting with a
responsive interface but also negotiating with a system that appears to
participate in the making, evaluation, and circulation of knowledge.

This study proposed the HAEPT, which offers an integrative structure for
explaining how that negotiation unfolds through three linked contracts:
epistemic, agency, and accountability. This is important because many of
the field's current concerns map onto one of these contracts but have
rarely been theorized together. The \emph{Epistemic Contract} addresses
trust, justification, verification, and the risk of accepting fluent
output as authoritative knowledge. The \emph{Agency Contract} addresses
how cognitive and metacognitive work is distributed between humans and
AI, which helps explain why performance gains do not necessarily
translate into learning gains (OECD, 2023). The \emph{Accountability
Contract} brings into the same frame issues that are often treated
separately---authorship, disclosure, academic integrity, fairness, and
ethical responsibility. Together, these contracts extend distributed
cognition and sociocultural perspectives by showing not only that
cognition is mediated across people and tools, but that GenAI also
reshapes the normative organization of knowledge work in educational
settings (Hutchins, 1995; Salomon, 1997). This is where we see HAEPT
adding something distinct: it does not replace prior theories but makes
visible a layer of educational user experience that those theories only
partially capture.

One reason GenAI research can appear contradictory is that users often
report both satisfaction and skepticism, or both trust and unease, in
the same study. HAEPT helps explain this not as an inconsistency, but as
an expected feature of human--AI epistemic partnership. Users
recalibrate over time. They test outputs, accept some, reject others,
and gradually adjust the extent of authority, agency, and responsibility
they assign to the system. This account builds on research on trust
calibration and trust in automation (Lee \& See, 2004) and aligns with
emerging measurement work showing that epistemic trust in GenAI is
multidimensional rather than a single attitude (Pandey et al., 2025).
The idea of partnership modes adds a useful middle layer between
isolated interactions and broad outcomes. It allows researchers to
describe recurrent configurations in ways that are theoretically
meaningful and empirically testable. In that sense, HAEPT not only
explains why user experiences differ but also how those differences are
patterned.

This paper also adds to the broader human--AI interaction literature by
showing why education cannot simply borrow human--AI teaming models
without adaptation. Human--AI teaming research has emphasized
transparency, coordination, situation awareness, and the active
management of trust over time (Endsley, 2023; McGrath et al., 2025;
Schmutz et al., 2024). These insights are highly relevant, but education
adds distinct demands. In education practice, the goal is not only
effective task completion but also the development of reasoning,
judgment, disciplinary understanding, and learner agency. Teachers
remain asymmetrically accountable for assessment, fairness, and the
credibility of credentials, even when students use AI independently (Gao
et al., 2025; UNESCO, 2023). HAEPT contributes here by showing that
educational human--AI interaction is a developmental and institutional
case, not merely a collaborative one. This matters because a system that
improves immediate performance may still be educationally weak if it
reduces students' epistemic ownership or encourages uncritical
dependence.

The two vignettes, CLAIS and ArgueAgent, demonstrate clear user
experience pictures that are distinct from conventional educational
technologies. CLAIS shows a partnership mode in which GenAI is
positioned as a reliable peer contributor, producing structured
co-agency, smooth interaction, and relatively low epistemic friction,
but also the risk of over-trusting AI explanations and under-specifying
accountability (Lee et al., 2025). ArgueAgent, by contrast, exhibits a
mode in which GenAI destabilizes ideas to stimulate argumentation,
producing stronger epistemic engagement but also the risk that students
become dependent on AI to structure and sustain discourse (Kleiman et
al., 2025). The value of this comparison is not only descriptive but
also demonstrates that different GenAI designs do not simply create
``more'' or ``less'' engagement; they produce different configurations
of authority, agency, and responsibility, and these configurations
matter educationally.

In that sense, the paper also contributes methodologically to how the
field might study GenAI going forward. Research on educational GenAI
often reports phenomena such as over-reliance, hallucination risk,
cheating, bias, or loss of critical thinking. These are important, but
when treated as isolated problems, they can lead to fragmented
responses. HAEPT provides a more organized system: Over-reliance can be
understood as a shift in the Agency Contract; Uncritical trust can be
understood as instability in the Epistemic Contract; Ambiguity about
authorship, disclosure, or fairness can be understood as a weak or
contested Accountability Contract. This reframing is more than
terminological. It suggests that interventions should target the
specific contract that is under strain rather than offer generic advice
to ``use AI responsibly,'' a meaningful step for research, design, and
policy, because it moves the conversation from broad caution to more
precise diagnosis.

Current debates can sometimes swing between enthusiasm and alarm: GenAI
is framed either as a powerful learning companion or as a threat to
integrity and human thinking. HAEPT suggests a more nuanced view. GenAI
can indeed support explanation, feedback, creativity, and scientific
reasoning, as growing work in assessment and science education already
indicates (Cooper \& Tang, 2024; Zhai \& Krajcik, 2024). But these
benefits do not emerge automatically from access or adoption, but depend
on how epistemic authority is calibrated, how cognitive work is
distributed, and how accountability is made visible. Put differently,
the educational question is not whether GenAI belongs in education, but
what kinds of human--AI partnerships education should cultivate.

\section{Conclusions and future
directions}\label{conclusions-and-future-directions}

This paper suggests that user experience with GenAI in education is best
understood not as simple tool use, but as a form of human--AI epistemic
partnership. What users experience when they work with systems such as
ChatGPT is not only convenience, speed, or interactivity. They also
experience shifting boundaries of knowledge, thinking, and
responsibility. HAEPT captures this by proposing that GenAI use is
organized through the ongoing negotiation of three interrelated
contracts: what counts as credible knowledge, who is doing the cognitive
work, and who remains answerable for the result. This perspective helps
explain why the same system can feel empowering in one context, risky in
another, and deeply ambivalent in both. It also helps explain why
adoption alone tells us very little about educational value. A
partnership can be efficient but shallow, engaging but overly dependent,
or productive but normatively unstable. The quality of the partnership
matters as much as the presence of the technology.

Future work should examine not only what users say about AI, but also
what they actually do with it across time. Looking ahead, although the
field has already begun developing instruments for learner--GenAI
relationships and epistemic trust (Pandey et al., 2025; Shi, 2025),
these efforts remain relatively separate. HAEPT suggests the need for a
more integrated measurement architecture. That means combining
self-report measures with discourse data, revision histories, prompting
patterns, verification behavior, disclosure decisions, and classroom
observation. Such an approach would make it possible to identify
partnership modes empirically, trace how users move between them, and
examine which configurations are associated with stronger learning,
greater epistemic agency, or more ethical use.

Future education systems could make uncertainty more visible to
strengthen the Epistemic Contract, require learners to explain why they
accepted or rejected AI output to preserve Agency, or embed disclosure
prompts and collaborative logs to clarify Accountability. We argue that
if HAEPT is useful, it should support design choices that intentionally
reshape the three contracts. Classroom interventions could also test
when AI support should be intensified and when it should fade. This is
especially important for dialogic and argumentation-based systems. Work
on AI-supported scientific argumentation already suggests that design
choices can either open space for deeper reasoning or unintentionally
centralize control in the AI (Watts et al., 2025). The next step is to
study these tradeoffs deliberately rather than treating them as side
effects.

Future research should also examine how different institutional
arrangements---such as disclosure rules, assessment redesign, teacher
professional learning, or discipline-specific AI policies---shape
partnership modes in practice. The Accountability Contract is unlikely
to stabilize through classroom practice alone. Students and teachers
need clearer norms around disclosure, authorship, acceptable assistance,
and assessment design. Recent guidance has emphasized that institutions
are still catching up to the pace of GenAI development, and this lag
leaves users navigating uncertainty with uneven support (UNESCO, 2023).
This line of work is especially important because accountability
pressures are not experienced equally. Teachers often carry the burden
of policy interpretation and assessment legitimacy, while students bear
the risks of inconsistency and unclear expectations (Crippen et al.,
2026).

We believe that future studies should examine how calibration of
partnership modes unfolds over weeks, semesters, and years, and how that
process varies by discipline, age, prior AI knowledge, language
background, and access to support. Partnership with GenAI is not fixed;
it develops. Students and teachers learn from repeated interactions,
from failures, from institutional signals, and from each other. This
matters not only for learning outcomes, but also for fairness. A
partnership mode that appears productive for one group may be
inaccessible or risky for another. If GenAI is becoming part of the
educational infrastructure, then questions of equity, bias, and
differential opportunity must be built into the study of user experience
rather than added later as a separate concern.

In closing, we see HAEPT as an attempt to offer the field a more precise
language for a changing educational reality. GenAI is already altering
how learners draft ideas, how teachers plan work, how feedback is
generated, and how knowledge moves through classrooms. The challenge is
no longer to decide whether a human--AI partnership exists. It does.
Instead, the more urgent task is to understand which forms of
partnership are educationally worth building, which ones should be
resisted, and how pedagogy, design, and policy can help move users
toward more reflective, responsible, and intellectually generative forms
of working with AI.

\section{Acknowledgement}\label{acknowledgement}

The research reported here was supported by the National Science
Foundation under Grant No. 2101104 (PI Zhai) and the Institute of
Education Sciences, U.S. Department of Education, through Grant
R305C240010 (PI Zhai). The opinions expressed are those of the authors
and do not represent the views of the National Science Foundation, the
Institute of Education Sciences, or the U.S. Department of Education.


\section*{References}\label{references}

Bahroun, Z., Anane, C., Ahmed, V., \& Zacca, A. (2023). Transforming
education: A comprehensive review of generative artificial intelligence
in educational settings through bibliometric and content analysis.
\emph{Sustainability, 15}(17), 12983.

Bai, Y., \& Wang, S. (2025). Impact of generative AI interaction and
output quality on university students' learning outcomes: a
technology-mediated and motivation-driven approach. \emph{Scientific
Reports, 15}(1), 24054.
\url{https://www.nature.com/articles/s41598-025-08697-6.pdf}

Bender, E. M., Gebru, T., McMillan-Major, A., \& Shmitchell, S. (2021).
On the dangers of stochastic parrots: Can language models be too big?
Proceedings of the 2021 ACM conference on fairness, accountability, and
transparency, pp. 610--623.

Bereiter, C., \& Scardamalia, M. (1993). Surpassing ourselves. \emph{An
inquiry into the nature and implications of expertise. Chicago: Open
Court}.

Bouyzourn, K., \& Birch, A. (2025). What Shapes User Trust in ChatGPT? A
Mixed-Methods Study of User Attributes, Trust Dimensions, Task Context,
and Societal Perceptions among University Students. \emph{arXiv preprint
arXiv:2507.05046}.

Bower, M., Torrington, J., Lai, J. W., Petocz, P., \& Alfano, M. (2024).
How should we change teaching and assessment in response to increasingly
powerful generative Artificial Intelligence? Outcomes of the ChatGPT
teacher survey. \emph{Education and Information Technologies}, 1--37.

Bretag, T., Harper, R., Burton, M., Ellis, C., Newton, P., van
Haeringen, K., Saddiqui, S., \& Rozenberg, P. (2019). Contract cheating
and assessment design: exploring the relationship. \emph{Assessment \&
Evaluation in Higher Education, 44}(5), 676--691.

Chan, C. K. Y., \& Hu, W. (2023). Students' voices on generative AI:
Perceptions, benefits, and challenges in higher education.
\emph{International Journal of Educational Technology in Higher
Education, 20}(1), 43.

Chen, B. (2025). Beyond tools: Generative AI as epistemic infrastructure
in education. \emph{arXiv preprint arXiv:2504.06928}.

Chi, M. T. (2009). Active‐constructive‐interactive: A conceptual
framework for differentiating learning activities. \emph{Topics in
cognitive science, 1}(1), 73--105.

Coeckelbergh, M. (2026). AI and epistemic agency: How AI influences
belief revision and its normative implications. \emph{Social
Epistemology, 40}(1), 59--71.

Cooper, G., \& Tang, K.-S. (2024). Pixels and Pedagogy: Examining
Science Education Imagery by Generative Artificial Intelligence.
\emph{Journal of Science Education and Technology}, 1--13.

Cotton, D. R., Cotton, P. A., \& Shipway, J. R. (2024). Chatting and
cheating: Ensuring academic integrity in the era of ChatGPT.
\emph{Innovations in Education and Teaching International, 61}(2),
228--239.

Crippen, K. J., Zhai, X., \& Lee, O. (2026). Responsible and Ethical
Principles for the Practice of AI-Supported Science Education. In X.
Zhai \& K. J. Crippen (Eds.), \emph{Advancing AI in Science Education:
Envisioning Responsible and Ethical Practice}. Springer.

Davis, F. D., Bagozzi, R., \& Warshaw, P. (1989). Technology acceptance
model. \emph{J Manag Sci, 35}(8), 982--1003.

Deng, N., Liu, E. J., \& Zhai, X. (2025). Understanding university
students' use of generative AI: The roles of demographics and
personality traits. International Conference on Artificial Intelligence
in Education, pp. 281--293.

Dillenbourg, P. (1999). \emph{Collaborative learning: Cognitive and
computational approaches. advances in learning and instruction series}.
ERIC.

Dwivedi, Y. K., Kshetri, N., Hughes, L., Slade, E. L., Jeyaraj, A., Kar,
A. K., Baabdullah, A. M., Koohang, A., Raghavan, V., \& Ahuja, M.
(2023). Opinion Paper:``So what if ChatGPT wrote it?'' Multidisciplinary
perspectives on opportunities, challenges and implications of generative
conversational AI for research, practice and policy. \emph{International
journal of information management, 71}, 102642.

Eaton, S. E. (2023). Academic integrity in the age of artificial
intelligence.

Endsley, M. R. (2023). Supporting Human-AI Teams: Transparency,
explainability, and situation awareness. \emph{Computers in Human
Behavior, 140}, 107574.

Engeström, Y. (2001). Expansive learning at work: Toward an activity
theoretical reconceptualization. \emph{Journal of Education and Work,
14}(1), 133--156.

Floridi, L., Cowls, J., Beltrametti, M., Chatila, R., Chazerand, P.,
Dignum, V., Luetge, C., Madelin, R., Pagallo, U., \& Rossi, F. (2018).
AI4People---An ethical framework for a good AI society: Opportunities,
risks, principles, and recommendations. \emph{Minds and machines,
28}(4), 689--707.
\url{https://iris.unito.it/bitstream/2318/1728327/1/Article_AI4PeopleAnEthicalFrameworkFor-2.pdf}

Gao, Y., Zhai, X., Li, M., Lee, G., \& Liu, X. (2025). A Multimodal
Interactive Framework for Science Assessment in the Era of Generative
Artificial Intelligence. \emph{Journal of Research in Science Teaching}.
\url{https://doi.org/10.1002/tea.70009}

Gee, J. P. (2000). Chapter 3: Identity as an analytic lens for research
in education. \emph{Review of research in education, 25}(1), 99--125.

Guo, S., Latif, E., Zhou, Y., Huang, X., \& Zhai, X. (2024a). Using
generative AI and multi-agents to provide automatic feedback.
\emph{arXiv preprint arXiv:2411.07407}.

Guo, S., Shi, L., \& Zhai, X. (2024b). Developing and validating an
instrument for teachers\textquotesingle{} acceptance of artificial
intelligence in education. \emph{Education and Information Technologies,
30}, 13439--13461.

Guo, S., Wang, Y., Yu, J., Wu, X., Ayik, B., Watts, F. M., Latif, E.,
Liu, N., Liu, L., \& Zhai, X. (2025). Artificial intelligence bias on
English language learners in automatic scoring. International Conference
on Artificial Intelligence in Education, pp. 268--275.

Herdiska, A., \& Zhai, X. (2024). Artificial Intelligence-Based
Scientific Inquiry. In X. Zhai \& J. Krajcik (Eds.), \emph{Uses of
Artificial Intelligence in STEM Education} (pp. 179--197). Oxford
University Press.

Heung, Y. M. E., \& Chiu, T. K. F. (2025, 2025/06/01/). How ChatGPT
impacts student engagement from a systematic review and meta-analysis
study. \emph{Computers and Education: Artificial Intelligence, 8},
100361.
\url{https://www.sciencedirect.com/science/article/pii/S2666920X25000013}

\url{https://www.sciencedirect.com/science/article/pii/S2666920X25000013?via\%3Dihub}

Holland, A., \& Ciachir, C. (2025). A qualitative study of students'
lived experience and perceptions of using ChatGPT: Immediacy, equity and
integrity. \emph{Interactive Learning Environments, 33}(1), 483--494.

Hutchins, E. (1995). \emph{Cognition in the Wild}. MIT press.

Januszewski, A., \& Molenda, M. (2013). \emph{Educational technology: A
definition with commentary}. Routledge.

Jin, S.-H. (2025). Measures of learner-generative ai relationships.
\emph{Computers and Education Open, 8}, 100258.

Johri, A., Hingle, A., \& Schleiss, J. (2024). Misconceptions,
pragmatism, and value tensions: Evaluating students\textquotesingle{}
understanding and perception of generative AI for education. 2024 IEEE
Frontiers in Education Conference (FIE), pp. 1--9.

Jose, B., Cleetus, A., Joseph, B., Joseph, L., Jose, B., \& John, A. K.
(2025). Epistemic authority and generative AI in learning spaces:
rethinking knowledge in the algorithmic age. Frontiers in Education,
Vol. 10, pp. 1647687.

Kasneci, E., Seßler, K., Küchemann, S., Bannert, M., Dementieva, D.,
Fischer, F., Gasser, U., Groh, G., Günnemann, S., \& Hüllermeier, E.
(2023). ChatGPT for good? On opportunities and challenges of large
language models for education. \emph{Learning and Individual
differences, 103}, 102274.

Kleiman, J., Gao, Y., Zhu, Z., Wang, Z., Zhou, Y., Hao, J., \& Zhai, X.
(2025, November, March 21, 2026). \emph{ArguAgent: A dialogue system for
scaffolding argumentative discourse} National Conference on AI in
Education, Athens, GA, United States)
\url{https://ai4genius.org/arguagent-poster/}

Kofinas, A. K., Tsay, C. H. H., \& Pike, D. (2025). The impact of
generative AI on academic integrity of authentic assessments within a
higher education context. \emph{British journal of educational
technology, 56}(6), 2522--2549.

Lee, G.-G., Mun, S., Shin, M.-K., \& Zhai, X. (2023). Collaborative
Learning with Artificial Intelligence Speakers (CLAIS): Pre-Service
Elementary Science Teachers\textquotesingle{} Responses to the
Prototype. \emph{Science \& Education}, 1--29.

Lee, J. D., \& See, K. A. (2004). Trust in automation: Designing for
appropriate reliance. \emph{Human factors, 46}(1), 50--80.

Lin, L., \& Pew Research Center. (2024). \emph{A quarter of U.S.
teachers say AI tools do more harm than good in K-12 education}.
\url{https://www.pewresearch.org/short-reads/2024/05/15/a-quarter-of-u-s-teachers-say-ai-tools-do-more-harm-than-good-in-k-12-education/}

Lund, B. D., Warren, S. J., \& Teel, Z. A. (2026). Measuring University
Students Satisfaction with Traditional Search Engines and Generative AI
Tools as Information Sources. \emph{arXiv preprint arXiv:2601.00493}.

Lyu, W., Zhang, S., Chung, T., Sun, Y., \& Zhang, Y. (2025).
Understanding the practices, perceptions, and (dis) trust of generative
AI among instructors: A mixed-methods study in the US higher education.
\emph{Computers and Education: Artificial Intelligence, 8}, 100383.

McCabe, D. L., Butterfield, K. D., \& Trevino, L. K. (2012).
\emph{Cheating in college: Why students do it and what educators can do
about it}. JHU Press.

McGrath, M. J., Duenser, A., Lacey, J., \& Paris, C. (2025).
Collaborative human-AI trust (CHAI-T): A process framework for active
management of trust in human-AI collaboration. \emph{Computers in Human
Behavior: Artificial Humans}, 100200.

Memarian, B., \& Doleck, T. (2023). ChatGPT in education: Methods,
potentials, and limitations. \emph{Computers in Human Behavior:
Artificial Humans, 1}(2), 100022.

Nazaretsky, T., Cukurova, M., \& Alexandron, G. (2021). An Instrument
for Measuring Teachers' Trust in AI-Based Educational Technology.

Ng, J., Tong, M., Tsang, E. Y., Chu, K., \& Tsang, W. (2025). Exploring
students' perceptions and satisfaction of using GenAI-ChatGPT tools for
learning in higher education: A mixed methods study. \emph{SN Computer
Science, 6}(5), 476.

Nyaaba, M., Shi, L., Nabang, M., Zhai, X., Kyeremeh, P., Ayoberd, S. A.,
\& Akanzire, B. N. (2024). Generative AI as a Learning Buddy and
Teaching Assistant: Pre-service Teachers\textquotesingle{} Uses and
Attitudes. \emph{arXiv preprint arXiv:2407.11983}.

OECD. (2023). \emph{OECD Digital Education Outlook 2023: Towards an
Effective Digital Education Ecosystem}. O. Publishing.
\url{https://doi.org/10.1787/c74f03de-en}

OECD. (2026). \emph{OECD Digital Education Outlook 2026: Exploring
Effective Uses of Generative AI in Education}. O. Publishing.

Pandey, C. S., Mishra, P., Pandey, S. R., \& Pandey, S. (2025).
Epistemic trust in generative AI for higher education scale (ETGAI-HE
scale). \emph{Ai \& Society}, 1--14.

Ravšelj, D., Keržič, D., Tomaževič, N., Umek, L., Brezovar, N., Iahad,
N. A., Abdulla, A. A., Akopyan, A., Segura, M. W. A., \& AlHumaid, J.
(2025). Higher education students' perceptions of ChatGPT: A global
study of early reactions. \emph{PloS one, 20}(2), e0315011.
\url{https://journals.plos.org/plosone/article/file?id=10.1371/journal.pone.0315011&type=printable}

Ryan, R. M., \& Deci, E. L. (2000). Self-determination theory and the
facilitation of intrinsic motivation, social development, and
well-being. \emph{American Psychologist, 55}(1), 68.

Sah, R., Hagemaster, C., Adhikari, A., Lee, A., \& Sun, N. (2025).
Generative AI in higher education: student and faculty perspectives on
use, ethics, and impact. \emph{Issues in Information Systems, 26}(2).

Salomon, G. (1997). \emph{Distributed cognitions: Psychological and
educational considerations}. Cambridge University Press.

Samala, A. D., Zhai, X., Aoki, K., Bojic, L., \& Zikic, S. (2024,
01/25). An In-Depth Review of ChatGPT's Pros and Cons for Learning and
Teaching in Education. \emph{International Journal of Interactive Mobile
Technologies (iJIM), 18}(02), pp. 96--117.
\url{https://online-journals.org/index.php/i-jim/article/view/46509}

\url{https://online-journals.org/index.php/i-jim/article/download/46509/14523}

Schmutz, J. B., Outland, N., Kerstan, S., Georganta, E., \& Ulfert,
A.-S. (2024). AI-teaming: Redefining collaboration in the digital era.
\emph{Current Opinion in Psychology, 58}, 101837.

Selwyn, N. (2021). \emph{Education and technology: Key issues and
debates}. Bloomsbury Publishing.

Shi, L. (2025). Assessing teachers' generative artificial intelligence
competencies: Instrument development and validation. \emph{Education and
Information Technologies}, 1--20.

Sweller, J. (1988). Cognitive load during problem solving: Effects on
learning. \emph{Cognitive science, 12}(2), 257--285.

Tsiani, M., Lefkos, I., \& Fachantidis, N. (2025). Perceptions of
generative AI in education: Insights from undergraduate and
master's-level future teachers. \emph{Journal of Pedagogical Research,
9}(2), 89--108.

UNESCO. (2023). \emph{Guidance for generative AI in education and
research}. S. a. C. O. United Nations Educational.

Venkatesh, V., Morris, M. G., Davis, G. B., \& Davis, F. D. (2003). User
acceptance of information technology: Toward a unified view1. \emph{MIS
Quarterly, 27}(3), 425--478.

Wang, Y., Wu, X., Huang, J., Liu, L., Zhai, X., \& Liu, N. (2026).
BRIDGE the Gap: Mitigating Bias Amplification in Automated Scoring of
English Language Learners via Inter-group Data Augmentation. \emph{arXiv
preprint arXiv:2602.23580}.

Waqas, M., Hasan, S., \& Ali, A. (2025). Effectiveness of Generative AI
Tools in Computer Science and Engineering Education. \emph{International
Journal of Artificial Intelligence in Education}, 1--32.

Watts, F. M., Liu, L., Ober, T. M., Song, Y., Jusino-Del Valle, E.,
Zhai, X., Wang, Y., \& Liu, N. (2025). A Framework for Designing an AI
Chatbot to Support Scientific Argumentation. \emph{Education Sciences,
15}(11), 1507.

Wertsch, J. V. (1998). \emph{Mind as action}. Oxford university press.

Zhai, X. (2022). ChatGPT user experience: Implications for education.
\emph{SSRN Electronic Journal}, 1--13.

Zhai, X. (2024). Transforming Teachers' Roles and Agencies in the Era of
Generative AI: Perceptions, Acceptance, Knowledge, and Practices.
\emph{Journal of Science Education and Technology}, 1--11.

Zhai, X. (2025). DAIL: Discipline-Based Artificial Intelligence
Literacy. \emph{Available at SSRN 5745703}.

Zhai, X., \& Krajcik, J. (2024). Artificial Intelligence-based STEM
Education. In X. Zhai \& J. Krajcik (Eds.), \emph{Uses of Artificial
Intelligence in STEM Education} (pp. 3--14). Oxford University Press.

Zhai, X., \& Nehm, R. (2023). AI and formative assessment: The train has
left the station. \emph{Journal of Research in Science Teaching, 60}(6),
1390--1398.

Zhai, X., Pellegrino, J. W., Rojas, M., Park, J., Nyaaba, M., Cohn, C.,
\& Biswas, G. (2026). Science Literacy: Generative AI as Enabler of
Coherence in the Teaching, Learning, and Assessment of Scientific
Knowledge and Reasoning. \emph{arXiv preprint arXiv:2603.06659}.



\end{document}